\documentstyle[floats,aps]{revtex}
\advance\topmargin 30pt
\def\subtilde#1{#1\llap{\lower2ex\hbox{$\widetilde{\hphantom{#1}}$}}}
\begin{document}
\input psfig
\pssilent
\title{Does loop quantum gravity imply $\Lambda=0$?}

\author{Rodolfo Gambini$^{1}$\footnote{Associate member of ICTP.}, 
Jorge Pullin$^2$}
\address{1. Instituto de F\'{\i}sica, Facultad de Ciencias, 
Tristan Narvaja 1674, Montevideo, Uruguay}
\address{
2. Center for Gravitational Physics and Geometry, Department of
Physics,\\
The Pennsylvania State University, 
104 Davey Lab, University Park, PA 16802}
\date{Mar 19th, 1998}
\maketitle
\begin{abstract}
We suggest that in a recently proposed framework for quantum gravity,
where Vassiliev invariants span the the space of states, the latter is
dramatically reduced if one has a non-vanishing cosmological
constant. This naturally suggests  that the initial state of the
universe should have been one with $\Lambda=0$.
\end{abstract}

\vspace{-6.5cm} 
\begin{flushright}
\baselineskip=15pt
CGPG-98/3-5  \\
gr-qc/9803097\\
\end{flushright}
\vspace{7cm}

For a long time the problem of the value of the cosmological constant
has been viewed as an important open issue in cosmology. The observed
value of the constant is quite low (smaller than $10\sp{-120}$ in
Planck units), and most field-theoretic mechanisms for generation of a
cosmological constant predict a much higher value. Although one can
set the constant to zero by hand, this amounts to a highly artificial
fine-tuning. It has therefore been viewed as desirable to find a
fundamental mechanism that could set its value to either zero or a
very small value. Quantum gravity has been suggested in the past as a
possible mechanism. The original proposal is due to Coleman \cite{Co},
and a recent attractive revision of this proposal was introduced by
Carlip \cite{Ca}. In both proposals the discussion was made at the
level of the path integral formulation of quantum gravity.  There have
also been proposals for ``screening'' of the cosmological constant in 
perturbative approaches to quantum gravity \cite{Woo}. Here we
would like to suggest that similar conclusions can be reached when
constructing the canonical quantum theory of general relativity.

Canonical quantum gravity was considered for a long time as an
intellectual wasteland. The complexity of the constraint equations
barred the implementation of even the earlier steps of a canonical
quantization program. This made the status whole canonical approach
look quite naive, since it could not even begin to grasp with the
physics of quantum gravity.  The general situation of the canonical
approach to quantum gravity started to get better with the
introduction of the Ashtekar new variables \cite{As86}; but still
important difficulties remain before a canonical quantization can be
completed. More specifically, the issues of observables and the
``problem of time'' are still important obstacles to the completion of
the canonical quantization program. In spite of this, it has become
recently clear that one is in fact able to make use of the formalism
to come up with interesting physical predictions, in spite of it being
incomplete. This is perhaps better displayed by the recent
calculations of black hole entropy \cite{Sm,Kr,Ro,AsBaCoKr}, which
yield physical predictions in spite of not addressing the issues
mentioned above. The intention of this note is to show that one can,
within the canonical approach, reach certain conclusions about the
value of the cosmological constant, again without having a complete
formalism. Contrary to the arguments about black hole entropy, those
that involve the cosmological constant require, as we will see, a
somewhat detailed use of the dynamics of the theory, namely the
Hamiltonian constraint. Unfortunately, although progress towards a
consistent and physically meaningful definition of the quantum
Hamiltonian constraint exists, the issue is not still settled. Our
calculations can therefore only be considered as preliminary. We will
make use of a recently introduced formulation of the Hamiltonian
constraint, that although appealing, has not yet been proved to be
completely consistent, especially at the level of the constraint
algebra.

Since the early days of the Ashtekar formulation, it became apparent
that the Hamiltonian constraint could be written in terms of a
differential operator in loop space called the loop derivative
\cite{Ga,BrPu93}. Writing the constraint in this way allowed to operate
explicitly on functions of loops and several solutions related to knot
invariants were found \cite{BrGaPuprl,BrGaPunpb,BrGaPuessay,GaPuBa} at
a formal level. This early work suffered from several drawbacks. On
one hand, it was difficult to consider functions of loops that were
compatible with the complicated Mandelstam identities \cite{GaPubook}
in loop space. Moreover, the loop derivative was not really well
defined if the wavefunctions were invariant under diffeomorphisms,
which is the case of interest in quantum gravity. Finally, the
Hamiltonian was regularized using a background metric, which
complicated reproducing the classical Poisson algebra of constraints
\cite{GaGaPu}. A recent series of results have improved the situation
regarding some of these issues. By working in the language of spin
networks, one is able to do away with the problem of the Mandelstam
identities \cite{RoSmvo}. Moreover, a well defined
prescription for the action of the loop derivative can be found
\cite{GaGrPu98}, and the knot polynomials that were formal solutions
can be constructed \cite{GaGrPu97} generalizing earlier work of Witten
and Martin \cite{WiMa}, and actually found to be solutions
of the constraints \cite{GaGrPu98}. More precisely, it was found
\cite{GaGrPu98,FrGaPu98}, that the knot invariants that are loop
differentiable are Vassiliev invariants, and in particular that the
Vassiliev invariants that arise from Chern--Simons theory are a
natural ``arena'' for the discussion of quantum gravity. An intuitive
way of understanding this is the fact that this set of invariants is
conjectured to separate all knots, and therefore is a natural basis in
the space of diffeomorphism invariant functions of loops.

We want here to discuss the construction of solutions with a
cosmological constant. We will assume our solutions to be of the form,
\begin{equation}
\Psi(\Gamma) = \sum_{n=p}^\infty v_{n}(\Gamma) \Lambda^{n-p},\qquad p>0
\label{sollambda}
\end{equation}
where $\Gamma$ is a spin network, $v$'s are Vassiliev invariants. We
have assumed here that solutions do not have essential singularities
in terms of $\Lambda$ (they could have negative powers of $\Lambda$
and one could reach the above expression by multiplying the
wavefunction times the appropriate powers of $\Lambda$). It can be
straightforwardly seen that $v_n(\Gamma)$ is actually a Vassiliev
invariant of order $n$. To see this, one should note that we are
interested in finding states that are annihilated by the Hamiltonian
constraint, 
\begin{equation}
H_\Lambda =H_0 +\Lambda {\rm det} g
\end{equation}
where $H_0$ is the vacuum (no cosmological constant) Hamiltonian and
the other term is proportional to the determinant of the three
metric. The quantum action of these operators \cite{GaGrPu98} is such
that $\hat{H}_0$ acting on a Vassiliev invariant of order $n$ produces
one of order $n-1$ and $\widehat{{\rm det} g}$ does not change the
order of the invariant it acts upon. Therefore the only chance of
cancellations to occur is if $v_n$ is an invariant of definite order,
$n$. 

It is straightforward to see from the above construction, taking the
limit $\Lambda\rightarrow 0$, that the leading term in each of the
expansions (\ref{sollambda}), $v_{p}$, has to be a solution to the
Hamiltonian constraint without cosmological constant, $\hat{H}_0$. In
particular, this reinforces the previous point. We know that solutions
of the Hamiltonian without cosmological constant are Vassiliev
invariants of a given order \cite{GaGrPu98}. We therefore have that
per each solution without cosmological constant, we could construct
one\footnote{Notice that the recurrence relation defines $v_n$ up to
solutions of $H_0$. Considering a given $v_k$ obtained by adding a
solution of $H_0$ is tantamount to superposing to the original chain a
new chain, starting with $v_k$ that corresponds to a new solution with
cosmological constant. Given that this is a linear space, it is always
possible to superpose solutions, but in keeping track of the number of
solutions one must count each solution only once.}  solution with
cosmological constant via the above procedure, i.e. constructing a
recurrence relation,
\begin{equation}
\hat{H}_0 v_n = \widehat{{\rm det} g} v_{n-1}.
\end{equation}

Up to here, therefore, it appears there are no less states with
cosmological constant than without, in fact, the number appears to be
the same. However, we have omitted an important requirement of these
quantum states, namely that they be diffeomorphism
invariants. Vassiliev invariants are diffeomorphism invariant
functions of loops, with the exception of the first order invariant
$v_1$, which is framing-dependent (it is closely related to the
self-linking number of a loop). That is, it is an invariant that
changes its values when one eliminates a twist from the loop, very
much as if it were an invariant of a ribbon rather than of a loop. In
the above construction, when $\hat{H}_0$ acts on a $v_n$ it will
generically produce as a result an invariant of order $n-1$ that
involves sums of products of primitive Vassiliev invariants of all
lower orders, including $v_1$ (for reasons of brevity, we omit the
detailed calculations here, we refer the reader to \cite{GaGrPu98}
for details). Therefore the result is not 
diffeomorphism invariant. We actually know of examples of solutions
like these, constructed by considering the loop transform of the
Chern--Simons state. One example of such solutions is the Kauffman
bracket knot polynomial, which is of the form (\ref{sollambda}), but
involves $v_1$ and powers of it at all orders in the expansion in
$\Lambda$.

\begin{figure}[t]
\centerline{\psfig{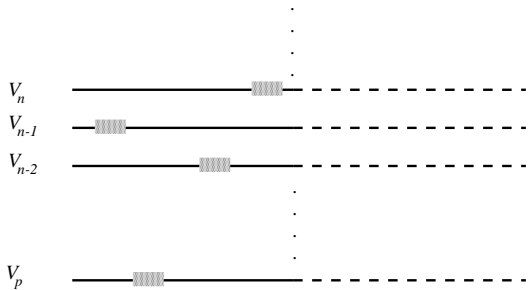}}
\label{fig1}
\caption{The restrictions that occur at each order of recurrence when
one requires diffeomorphism invariance.}
\end{figure}

Therefore, if we wish to require that the resulting wavefunction be
diffeomorphism invariant, this will place restrictions on our
constructing procedure. This is illustrated in figure \ref{fig1}.  The
figure represents pictorially by a horizontal line the set of all
Vassiliev invariants of a given order. The solid line represents those
that are constructed not involving $v_1$ and the dashed line those
that do. The figure is roughly up to scale, the fraction of invariants
that does not depend on $v_1$ depends on the number of new primitive
Vassiliev invariants that appear at each order. This number is not
known precisely, but it is conjectured to grow exponentially. One can then
make an estimate of the fraction of invariants dependent on $v_1$ to be
at least $1/3$ of all invariants.  Consider the lower $n$ 
examples,
\begin{eqnarray}
V_1:&&\qquad v_1\\
V_2:&&\qquad v_1^2, v_2\\
V_3:&&\qquad v_1^3, v_2 v_1, v_3\\
V_4:&&\qquad v^{(1)}_4, v^{(2)}_4, v_3 v_1, v_2 v_1^2, v_2^2, v_1^4,
\end{eqnarray}
the superscripts on $v_4$  denote that there are more than one primitive
Vassiliev invariants of order four. 

Now, when the Hamiltonian acts on elements of level $V_n$ that do not
contain $v_1$, it will generically produce an invariant of $V_{n-1}$
that contains $v_1$. In this case the original element should be
discarded as a potential candidate to generate a diffeomorphism
invariant state. The same argument repeats itself at each level. The
end result is that in order for the recursion relation to yield
framing independent invariants at all orders, one is requiring a very
stringent condition on the initial invariant of the series $v_p$. That
invariant had, to begin with, to be a solution of the vacuum
Hamiltonian $H_0$. In addition to that, its has to belong in the
diffeo-invariant image of the diffeo-invariant portion of $V_1$
through the action of $H_0$, and so on for each order. The recursion
relation has to ``thread the hoops'' determined by the shaded areas of
figure 1. At the very least this implies that there are less solutions
of the Hamiltonian constraint starting at a given order $v_p$ with
cosmological constant than without, if one imposes diffeomorphism
invariance. In fact, it appears there could be significantly less. The
solutions of the type (\ref{sollambda}) cannot be constructed with
finite series (it would require $\widehat{\rm det g}$ to vanish on the
higher order invariant for all spin networks). Therefore one is
imposing an infinite number of additional conditions to be satisfied. 
One cannot make more precise these arguments, since one is dealing
with infinite dimensional spaces, but it appears that the requirement
is very stringent. It is a fact that we do not know a single solution
satisfying this requirement up to now.

We have therefore shown that it is plausible that the canonical theory
of quantum general relativity with cosmological constant includes a
significantly smaller number of states than vacuum general relativity.
In particular, it could be that there are {\em no} solutions with 
cosmological constant. In the latter case, it clearly would prevent us
from having a cosmological constant in nature. Even if some solutions
with cosmological constant do exist, the fact that gravity with
cosmological constant would represent  a smaller portion of ``phase
space'' than vacuum general relativity, would make $\Lambda=0$ a more
statistically preferred possibility. This could be relevant in
settings in which the cosmological constant is allowed to vary
primordially, be it through couplings to matter or via ``natural
selection'' mechanisms like proposed in \cite{Smbook}. 

It is appropriate to finish by pointing out the weak points of the
argument. To begin with, the setting in which we are discussing
quantum gravity is not completely established. The Hamiltonian
constraint we are considering is the ``doubly densitized'' constraint,
that cannot be easily promoted to a regularization-independent quantum
operator. This will inevitably bring problems when one wants to check
its quantum Poisson algebra. Although we know a few solutions of this
Hamiltonian constraint, there is still a lot to be learnt about its
space of solutions, especially in the context of spin networks. To be
fair, it is also true that we are not using an enormous amount of
detail about the action of the constraint, we basically only need that
the action on a Vassiliev invariant of order $n$ give as result
invariants of order up to $n-1$ in order to construct the
argument. Many candidates for Hamiltonians are expected to exhibit
this generic property. For our argument to be turned into a rigorous
one, one would need a much better control of the space of solutions in
order to introduce some sort of measure to determine ``how small'' is
the space of solutions with cosmological constant. The main purpose of
this note was not to provide a rigorous argument, but to suggest that
in the context of canonical quantization issues like that of the value
of the cosmological constant can be addressed. And that with the
better control we are gaining on the theory at this level, might be
settled rigorously in the near future.

We wish to thank Laurent Freidel, Jorge Griego and Charles Torre for
discussions.  This work was supported in part by grants
NSF-INT-9406269, NSF-PHY-9423950, research funds of the Pennsylvania
State University, the Eberly Family research fund at PSU and PSU's
Office for Minority Faculty development. JP acknowledges support of
the Alfred P. Sloan foundation through a fellowship. We acknowledge
support of Conicyt (project 49) and PEDECIBA (Uruguay).

\end{document}